# Contrasting magnetoelectric behavior in multiferroic hexaferrites as understood by crystal symmetry analyses

Y. S. Chai,[1,2,3*] S. H. Chun,[1] J. Z. Cong,[2] and Kee Hoon Kim[1*]

[1] *Center for Novel States of Complex Materials and Research (CeNSCMR) and Institute of Applied Physics, Department of physics and astronomy, Seoul National University, Seoul 151-747, Republic of Korea*
[2] *Beijing National Laboratory for Condensed Matter Physics, Institute of Physics, Chinese Academy of Sciences, Beijing 100190, China*
[3] *Department of Applied Physics, Chongqing University, Chongqing, 401331, China*

*Correspondence and requests for materials should be addressed to K. H. Kim (optopia@snu.ac.kr) and Y. S. Chai (hawkchai@gmail.com)

**Magnetoelectric (ME) properties under rotating magnetic field *H* are comparatively investigated in two representative hexaferrites Y-type $Ba_{0.5}Sr_{1.5}Zn_2(Fe_{0.92}Al_{0.08})_{12}O_{22}$ and Z-type $Ba_{0.52}Sr_{2.48}Co_2Fe_{24}O_{41}$, both of which have exhibited a similar transverse conical spin structure and giant ME coupling near room temperature. When the external *H* is rotated clockwise by 2π, in-plane *P* vector is rotated clockwise by 2π in the Y-type hexaferrite and counterclockwise by 4π in the Z-type hexaferrite. A symmetry-based analysis reveals that the faster and opposite rotation of *P* vector in the Z-type hexaferrite is associated with the existence of a mirror plane perpendicular to *c*-axis. Moreover, such a peculiar crystal symmetry also results in contrasting microscopic origins for the spin-driven ferroelectricity; only the inverse DM interaction is responsible for the Y-type hexaferrite while additional *p-d* hybridization becomes more important in the Z-type hexaferrite. This work demonstrates the importance of the crystal symmetry in the determination of ME properties in the hexaferrites and provides a fundamental framework for understanding and applying the giant ME coupling in various ferrites with hexagonal crystal structure.**

**Introductions**

Multiferroics are materials that combine the magnetic and ferroelectric orders[1-3]. Their possible magnetoelectric (ME) effects—the response of electric polarization $P$ to a magnetic field ($H$) or magnetization to an electric field ($E$) —have attracted great attention because of the potential application for novel electronic devices[4-5]. In particular, the spin-driven multiferroics, of which ferroelectricity has a magnetic origin, have been a focus due to their strong and versatile ME effects[6]. Therefore, in terms of both fundamental and technological points of view, it will be necessary to understand the microscopic origin of ferroelectric polarization $P$ as well as the accompanied ME behaviors.

It is known that the microscopic origin of polarization and related ME responses are correlated with the magnetic ordering pattern or the associated magnetic symmetry[6]. On one hand, several microscopic ME mechanisms based on the magnetic ordering pattern have been proposed such as the spin-current (or inverse Dzyaloshinskii-Moriya (DM) interaction) $P \propto P^{DM} e_{12} \times (S_1 \times S_2)$[7,8], exchange-striction $P \propto P^{ES} e_{12}(S_1 \cdot S_2)$ where $e_{12}$ is connecting vector of adjacent spin pair $S_1$ and $S_2$ [9,10], and p-d hybridization $P \propto P^{pd} e_1(e_1 \cdot S_1)^2$ where vector $e_1$ connects the transition metal and its neighbor ligand atom[11-13]. On the other hand, direct relationship between the ME effect and a magnetic point group is established since the initial work on the magnetoelectric $Cr_2O_3$[14]. Nevertheless, the ME effects in pure magnetoelectrics is very weak while it is very strong in some of the spin-driven multiferroics because their spin order sensitively responds to applied $H$.

One of the highest ME coefficients (=$dP/dH$) in spin-driven multiferroics were found in hexaferrite systems both at low temperatures and at room temperature. The crystal structure of hexaferrites can be described as the stacking sequence of spinel-like S, tetragonal-like T, and rhombohedral like R blocks, *i.e.*, **S-block** ($Me^{2+}Fe_4O_8$; $Me^{2+}$= divalent metal ion), **T** [$(Ba,Sr)_2Fe_8O_{14}$] and **R** [$(Ba,Sr)Fe_6O_{11}$]$^{2-}$ structure blocks along [001] direction. According to the different sequences, hexaferrites are classified into six main types depending on their chemical formulas and stacking sequences: M-type [$(Ba,Sr)Fe_{12}O_{19}$], W-type [$(Ba,Sr)Me_2Fe_{16}O_{27}$], X-type [$(Ba,Sr)_2Me_2Fe_{28}O_{46}$], Y-type [$(Ba,Sr)_2Me_2Fe_{12}O_{22}$], Z-type [$(Ba,Sr)_3Me_2Fe_{24}O_{41}$], and U-type [$(Ba,Sr)_4Me_2Fe_{36}O_{60}$]. In particular, Y-type and Z-type have **STSTST** and **STSRSTSR** sequences, respectively (Fig. 1a)[15]. For every **S** and **T** block, there are space inversion centers in the middle while there is mirror plane ($\perp$[001]) for **R** block instead. Therefore, Y-type has a -3*m* point group while Z-type has a 6/*mmm* point group with the extra mirror plane. Both of them are centrosymmetric without ferroelectricity in high temperature. However, a so-called transverse cone spin configuration at low *T* and low *H* could break the space inversion center in those hexaferrites and induce the in-plane polarization. Due to the very sensitive response of transverse cone to the small external *H*, strong ME effects are found in M-type Ba(Fe,Sc)$_{12}$O$_{19}$, Y-type $(Ba,Sr)_2Me_2(Fe,Al)_{12}O_{22}$, Z-type $(Ba_{1-x}Sr_x)Co_2Fe_{24}O_{41}$, and U-type $Sr_4Co_2Fe_{36}O_{60}$ with transverse cone phase up to room *T*. The accompanied converse ME effects, i.e., electric field controlled large magnetization reversal or changes are also demonstrated in Y-type

Ba$_{0.5}$Sr$_{1.5}$Zn$_2$(Fe$_{0.92}$Al$_{0.08}$)$_{12}$O$_{22}$ and Z-type Ba$_{0.52}$Sr$_{2.48}$Co$_2$Fe$_{24}$O$_{41}$, respectively. However, their ME behaviors are different—direct in-plane $H$ reversal can fully reverse the $P$ value in Y-type, but not in Z-type when TC phases persist around zero field, leading to a giant ME effect at zero-$H$ for Y-type and a large ME effect at finite $H$ for Z-type.

To thoroughly understand the physical origin of the difference in their ME behaviors, the lattice symmetry has to be considered. The ferroelectric polarization arises from the reduced lattice symmetry and thereby its emergence is constrained by the original symmetry. However, this aspect has not been accounted for the description of ME characteristics based on the magnetic structure. We find that the original lattice symmetry in hexaferrite systems has a prominent role to determine dominance of ME behaviors originating from the possible microscopic mechanisms. In this work, distinct ME behaviors in the ME hexaferrites Y-type Ba$_{0.5}$Sr$_{1.5}$Zn$_2$(Fe$_{0.92}$Al$_{0.08}$)$_{12}$O$_{22}$ (BSZFAO) and Z-type Ba$_{0.52}$Sr$_{2.48}$Co$_2$Fe$_{24}$O$_{41}$ (BSCFO) are observed and attributed to the additional mirror plane in the crystal symmetry of Z-type, based on a local-symmetry-based theoretical analysis. The microscopic origins of dominant $P$ are found to be different between two systems due to the existence of additional mirror plane in Z. In particular, the dominant polarization of Z-type is induced by non-spin current mechanisms.

**Results**

**Magnetic and ME behaviors of Y-type and Z-type.** The Y-type and Z-type hexaferrites have very similar in-plane magnetic field induced commensurate transverse conical (TC) magnetic phase (Fig. 1)[16-18], but distinct crystal symmetries. To understand their magnetic structures, we will follow an approximation used in the previous studies[15-19]. All the spins are conveniently divided into alternating stacked $L$ blocks with large spin moments ($\mu_L$) and $S$ blocks having small spin moments ($\mu_S$) (Fig. 1), which are different from the **R**, **S**, and **T** structure blocks. In each magnetic block, all the spins are assumed to be parallel for simplicity. According to the previous neutron diffraction measurements around $H = 0$, Y-type hexaferrite at low temperature (< 100 K) after a high $H$ history and Z-type below 400 K have similar commensurate TC phase with propagation vector $k_0 = (0,0,3/2)$ and $k_0 = (0,0,1)$ respectively (Fig. 1). As can be seen in Fig. 1, both TC phases have exactly the same magnetic block periodicity and spin texture that $L$ blocks and $S$ blocks have antiferromagnetically ordered in-plane components and out-of-plane components, respectively, whereas they are antiparallel along the cone directions. In terms of the lattice symmetry of magnetic blocks, it is the same for their $S$ blocks but different for $L$ blocks in that there is a space inversion center in the $L$ blocks of Y-type while a mirror plane ($\perp[001]$) in that of Z-type.

Regardless of the differences in their crystal symmetries, previous studies found that the TC phases in both systems are ferroelectric with in-plane polarization[15,20]. It was widely believed that the microscopic origin of $P$ in all the TC phases was the spin-current mechanism: $P \propto \Sigma k_0 \times (\mu_L \times \mu_S)/|k_0|$. This mechanism will lead to a $P$ vector perpendicular to $k_0$ (//[001]) and cone axis (//$H$ for large $H$, as shown in Fig. 1a), very similar to the case of CoCr$_2$O$_4$[21]. To verify such orthogonal relationship between $P$, $H$ and $k_0$ direction, there are two methods, 1) the reversing of in-plane $H$ can lead to a reversal of the $P$ vector (Fig. 2a) if the spin helicity $\Sigma k_0 \times (\mu_L \times \mu_S)$ could be preserved on passing the $H = 0$ point[22]. 2) horizontally rotating $H$ can lead to the projection of $P$ vector along a fixed direction to show a sinusoidal behavior as the function of rotating angle ϕ (Fig. 2b)[22]. In some Y-type hexaferrites, the TC induced $P$ has passed the above two tests[15,22]. On the other hand, some recent in-situ X-ray diffraction studies on the Z-type hexaferrite imply a violation to these tests[23]. In this work, we performed similar experiments on both Y and Z-type hexaferrites.

We characterize the ME properties of the Y-type BSZFAO at 30 K where the TC phase is stabilized between ± 15 kOe. Figure 2c shows that the $P_y$ can be fully reversed by the reversal of perpendicular $H_x$, indicating a conservation of $P \perp H \perp k_0$ relationship during such process (Fig. 2a). Figure 2d demonstrate the angle ϕ dependent $P_y$(ϕ) under rotating $H$ of 0.2 and 2 kOe for BSZFAO. Here, ϕ is defined as the relative angle between $H$ and $x$-axis in the clockwise direction. It is very clear that $P_y$ shows nearly cosϕ dependent behaviors with small hysteresis for both field values, also indicating the orthogonal relationship between $P$, $H$ and $k_0$ directions in Fig. 2b. This behavior is essentially the same as that reported for other Y-type hexaferrites[18,22]. The above ME behaviors strongly suggests that the spin-current model holds for the TC induced $P$ in our Y-type hexaferrite.

On the contrary, the inverse of $P$ vector by $H$ reversal has never been reported in any Z-type hexaferrites with TC phase stabilized around $H = 0$[17,18,24]. We also verify this in our Z-type BSCFO at 305 K where the TC phase should be stabilized within the range of ± 20 kOe[18,24]. As shown in Fig. 2e, ±$P_y$ just quickly approach zero and recovers to a slightly smaller magnitude with the same sign after $H_x$ reversal, similar to the previous investigations. A possible explanation could be the effects from unknown magnetic phase or exotic domain structure around $H_x = 0$[25]. To eliminate these possibilities, we performed the angle ϕ dependent polarization measurement along $y$ direction under horizontally rotating $H$ of 5 and 12 kOe for BSCFO at 305 K. Here, ϕ is defined as the relative angle between $H$ and $x$-axis in the clockwise direction. These two magnetic fields are enough to keep the sample a single TC domain state following the $H$ direction during the rotation processes. Surprisingly, the $P_y$ show roughly the cos(2ϕ) dependent behaviors for both $H$ values. This novel behavior is in sharp contrast to the case of Y-type hexaferrites. Due to the large background signal in ME current at 305 K, the absolute $P_y$ value cannot be reliably estimated in these measurements. However, we can still conclude that the dominant cos(2ϕ) dependent $P_y$ in the TC phase of Z-type BSCFO strongly indicates a microscopic origin different from the expected spin-current model.

To map out the complete trajectory of $P$ vector under reversing and rotating $H$, we polished the sample in roughly cubic shape with two pairs of orthogonal electrodes along $x$ and $y$ directions, respectively, as shown in Fig. 3c. Both $P_x$ and $P_y$ can be monitored simultaneously in this configuration. The $P$ measurements were performed at 10 K to reduce the background signal and obtain the absolute $P$ values.

As shown in Figs. 4a and 4b, neither $P_x$ nor $P_y$ reverses its signs after $H_x$ reversal. The estimated directions of $P$ vector are almost the same for ±4 kOe, consistent with the results at 305 K. In addition, the magnitude of $P_y$ is much larger than that of $P_x$, which roughly keeps the $P \perp H$ ($\perp k_0$) relationship. Then, to further test the orthogonal relationship, we measured $P_x$ and $P_y$ simultaneously in horizontally rotating $H$ of 4 and 10 kOe (Figs. 3c-g). More surprisingly, the $P_x$ and $P_y$ show dominating $\sin(2\phi)$ and $\cos(2\phi)$ behaviors respectively, with a relative phase difference of about 45 degree. As a result, the calculated $P$ vector rotates in opposite direction and roughly twice faster in speed than that of the $H$ vector, as schematically illustrated in Fig. 3h. There may be some small $\sin\phi$ or $\cos\phi$ components which are close to our technical limitations. This ME behavior resembles the case of triangular-lattice helimagnet $MnI_2$, which exhibits the $P$ vector smoothly rotates clockwise twice while the $H$ vector rotates counterclockwise once at certain critical field region[26]. This feature has been interpreted in terms of $H$ switching of the multiple in-plane propagation vectors domains. However, BSCFO only hosts an out-of-plane $k_0 = (0,0,1)$ single domain state for the above in-plane $H$ values. Nevertheless, the ME behaviors and ferroelectricity in Z-type BSCFO are beyond the prediction of spin-current mechanism and must have other physical origins.

**Symmetry analysis of the electric dipole produced by two spins.** So far, no practical first-principle calculation approach is possible to resolve the microscopic origin of spin-induced ferroelectricity in any ME hexaferrite systems due to their extremely large unit cells and complex site-by-site spin structures. Instead, we will analyze their microscopic origin of *P* and ME behaviors by some recent developed local symmetry theories[27-29].

In general, the local electric-dipole *p* caused by a spin pair $\mu_1$ and $\mu_2$ or by one of the spins with its surroundings can be universally expressed as the quadratic functions in Einstein convention:

$$p_{ij}^{\gamma} = P_{ij}^{\alpha\beta\gamma}\mu_i^{\alpha}\mu_j^{\beta} \text{ and } p_{ii}^{\gamma} = P_{ii}^{\alpha\beta\gamma}\mu_i^{\alpha}\mu_i^{\beta} \tag{1}$$

where *α, β, γ* run over all the Cartesian coordinates, *x, y, z*; *i, j* run over the site or spin labels, *1,2*. $P_{ii}^{\alpha\beta\gamma}$ and $P_{ij}^{\alpha\beta\gamma}$ can be regarded as a kind of single-spin tensor and two-spin tensor, respectively, according to their definitions[28]. Then total polarization $P_{\text{tot}}$ of a multiferroics with spin-induced ferroelectricity can be concisely expressed by those local ME tensors as:

$$P_{\text{tot}}^{\gamma} \propto \sum_{i \neq j} p_{ij}^{\gamma} + \sum_i p_{ii}^{\gamma} = \sum_{i \neq j} P_{ij}^{\alpha\beta\gamma}\mu_i^{\alpha}\mu_j^{\beta} + \sum_i P_{ii}^{\alpha\beta\gamma}\mu_i^{\alpha}\mu_i^{\beta} \tag{2}$$

Where *α, β, γ* run over all the Cartesian coordinates, *x, y, z*; *i, j* run over the site or spin pair labels in a magnetic unit cell of the multiferroics. Note that the forms of the local ME tensors are site or site-pair-specific and dictated only by the lattice symmetry of the site or site-pair since they are third rank **polar** tensors [see the Supplementary Note 1]. The matrix form of each tensor can be mathematically transformed by symmetry operators and simplified according to the local crystal symmetries, see Supplementary Note 1 for detailed calculations. Therefore, the Eq. (2) of a multiferroics can also be simplified according to its crystal symmetry. Hexaferrite systems have very high lattice symmetry which will put severe symmetric constraints over the forms of their local ME tensors. Therefore, we may adopt this method in both hexaferrite systems to calculate the $P_{tot}$ with a much-simplified analytical form.

Moreover, all three known mechanisms are the special cases of the local ME single spin tensor $P_{ii}^{\alpha\beta\gamma}$ and two-spin tensor $P_{ij}^{\alpha\beta\gamma}$ [27,28]. In particular, the spin-current mechanism $P \propto P^{DM}e_{12}\times(\mu_1\times\mu_2)$ will allow a two-spin tensor $P_{ij}^{\alpha\beta\gamma}$ with antisymmetric matrix form:

$$P^{DM}\begin{pmatrix} 0 & e_{12y},-e_{12x},0 & e_{12z},0,-e_{12x} \\ -e_{12y},e_{12x},0 & 0 & 0,e_{12z},-e_{12y} \\ -e_{12z},0,e_{12x} & 0,-e_{12z},e_{12y} & 0 \end{pmatrix} \quad (3)$$

while exchange-striction $P \propto P^{ES}e_{12}(\mu_1\cdot\mu_2)$ allows a two-spin tensor $P_{ij}^{\alpha\beta\gamma}$ with non-zero diagonal form only:

$$P^{ES}\begin{pmatrix} e_{12x},e_{12y},e_{12z} & 0 & 0 \\ 0 & e_{12x},e_{12y},e_{12z} & 0 \\ 0 & 0 & e_{12x},e_{12y},e_{12z} \end{pmatrix} \quad (4)$$

where $e_{12} =(e_{12x}, e_{12y}, e_{12z})$ is the connecting vector of $\mu_1$ and $\mu_2$.
In contrast, p-d hybridization mechanism $P \propto P^{pd}e_i(e_i\cdot\mu_i)^2$ gives a single spin tensor $P_{ii}^{\alpha\beta\gamma}$ with the symmetric matrix form:

$$P^{pd}\begin{pmatrix} e_{ix}e_{ix}e_{ix},e_{ix}e_{ix}e_{iy},e_{ix}e_{ix}e_{iz} & e_{ix}e_{iy}e_{ix},e_{ix}e_{iy}e_{iy},e_{ix}e_{iy}e_{iz} & e_{ix}e_{iz}e_{ix},e_{ix}e_{iz}e_{iy},e_{ix}e_{iz}e_{iz} \\ e_{iy}e_{ix}e_{ix},e_{iy}e_{ix}e_{iy},e_{iy}e_{ix}e_{iz} & e_{iy}e_{iy}e_{ix},e_{iy}e_{iy}e_{iy},e_{iy}e_{iy}e_{iz} & e_{iy}e_{iz}e_{ix},e_{iy}e_{iz}e_{iy},e_{iy}e_{iz}e_{iz} \\ e_{iz}e_{ix}e_{ix},e_{iz}e_{ix}e_{iy},e_{iz}e_{ix}e_{iz} & e_{iz}e_{iy}e_{ix},e_{iz}e_{iy}e_{iy},e_{iz}e_{iy}e_{iz} & e_{iz}e_{iz}e_{ix},e_{iz}e_{iz}e_{iy},e_{iz}e_{iz}e_{iz} \end{pmatrix} \quad (5)$$

where vector $e_i = (e_{ix}, e_{iy}, e_{iz})$ connects the $\mu_i$ and its neighbor ligand atom ($i$ = 1,2). Therefore, from the non-zero components of the simplified $P_{tot}$, we could also deduce its microscopic origin by comparing it with Eqs. (3)-(5).

**Calculation of spin-driven polarization and related ME properties of both Hexaferrites.** To conveniently compare with the experimental observations, we will calculate the angle dependent $P_{tot}$ of both hexaferrites under in-plane rotating $H$ since the spin configuration is fixed to a single domain TC state. Instead of using the full atom-by-atom crystal and spin models, we first adopt two simplified generic models for the crystal structures of the Y-type and Z-type respectively, which preserve the crystal point groups in the paramagnetic phases with the essential symmetric operations (Fig. 4a). The lattices are divided according to the magnetic blocks instead of structure blocks, where each magnetic block is represented by two atomic layers connected by either a spatial inversion center or a mirror in the middle (Fig. 4). Each atomic layer consists of three identical transition metal ions forming an equilateral triangle (we don't consider the difference between Fe and Co ions to simplify the models). Moreover, a three-fold rotation along [001] direction and three mirrors including [001] axis are also allowed for each layer, and subsequently for these models, as shown in Figs. 1a, 1b and 4b.

Then, we assume that for the initial $H_x$, the initial TC configuration would have a spin configuration in one magnetic unit cell:

$\mu_{S1}=(-\mu_S^x, 0, -\mu_S^z)$,

$\mu_{S2}=(-\mu_S^x, 0, \mu_S^z)$

$\mu_{L1}=(\mu_L^x, -\mu_L^y, 0)$

$\mu_{L2}=(\mu_L^x, \mu_L^y, 0)$ (6)

Where $\mu_{S1}$, $\mu_{S2}$, $\mu_{L1}$ and $\mu_{L2}$ are the total spin vectors in $S_1$, $S_2$, $L_1$ and $L_2$ blocks within one magnetic unit cell respectively, the coefficients $\mu_S^x$, $\mu_S^z$, $\mu_L^x$ and $\mu_L^y$ are the initial spin components in $\mu_{S2}$ and $\mu_{L2}$ along three Cartesian coordinates. As to describe their rotating angle dependent spin patterns, we assume that TC is rigid and the cone axis follows precisely the in-plane $H$ direction:

$$\mu_{S1}(\phi) = (-\mu_S^x \cos\phi, \mu_S^x \sin\phi, -\mu_S^z),$$

$$\mu_{S2}(\phi) = (-\mu_S^x \cos\phi, \mu_S^x \sin\phi, \mu_S^z)$$

$$\mu_{L1}(\phi) = (\mu_L^x \cos\phi - \mu_L^y \sin\phi, -\mu_L^x \sin\phi - \mu_L^y \cos\phi, 0)$$

$$\mu_{L2}(\phi) = (\mu_L^x \cos\phi + \mu_L^y \sin\phi, -\mu_L^x \sin\phi + \mu_L^y \cos\phi, 0) \qquad (7)$$

where $\phi$ is the angle between $H$ and x-axis defined in Fig. 3c. Finally, we assume that all the six atoms in each block have exactly the same spin for simplicity, or 1/6 of total moment within a block. With the above lattice and spin configurations for both Y and Z-type hexaferrites, we could calculate their $\phi$ dependent $P_{tot}$ according to Eq. (2).

We find that the summation of single-spin tensor in one block $\sum_i P_{ii}^{\alpha\beta\gamma}$ has a simplified matrix form according to the lattice symmetries, as shown the Table 1 (See Supplementary Note 2 for detailed discussion). The matrix forms of $\sum_i P_{ii}^{\alpha\beta\gamma}$ in $S_2$ and $L_2$ can be calculated in that they are connected with $S_1$ and $L_1$ respectively by space inversion operator. In this case, only the $\sum_i P_{ii}^{\alpha\beta\gamma}$ of $L$ blocks in Z-type allow nonzero matrix components with one independent coefficient due to the existence of mirror $m\perp[001]$ in the $L$ blocks of Z-type. Every summation of $\sum_i P_{ii}^{\alpha\beta\gamma}$ in the blocks with the existence of a space inversion center are exactly zero.

Table 1. The simplified $\sum_i P_{ii}^{\alpha\beta\gamma}$ in $S_1$ and $L_1$ magnetic blocks for both hexaferrite systems.

|  | $S_1$ | $L_1$ |
|---|---|---|
| Y-type | $\begin{pmatrix} 0,0,0 & 0,0,0 & 0,0,0 \\ 0,0,0 & 0,0,0 & 0,0,0 \\ 0,0,0 & 0,0,0 & 0,0,0 \end{pmatrix}$ | $\begin{pmatrix} 0,0,0 & 0,0,0 & 0,0,0 \\ 0,0,0 & 0,0,0 & 0,0,0 \\ 0,0,0 & 0,0,0 & 0,0,0 \end{pmatrix}$ |
| Z-type | $\begin{pmatrix} 0,0,0 & 0,0,0 & 0,0,0 \\ 0,0,0 & 0,0,0 & 0,0,0 \\ 0,0,0 & 0,0,0 & 0,0,0 \end{pmatrix}$ | $6\begin{pmatrix} a_0,0,0 & 0,-a_0,0 & 0,0,0 \\ 0,-a_0,0 & -a_0,0,0 & 0,0,0 \\ 0,0,0 & 0,0,0 & 0,0,0 \end{pmatrix}$ |

For the summation of two-spin tensor $\sum_{ii'} P_{ii'}^{\alpha\beta\gamma}$, there are inter-block and intra-block cases, see Fig. 4c. Whatsoever, $\sum_{ii'} P_{ii'}^{\alpha\beta\gamma}$ have simplified matrix forms according to the lattice symmetries, as shown the Table 2 (see also the Supplementary Note 2). Other inter and intra-block tensor summations can be inferred accordingly via space inversion or mirror operation.

Table 2. The simplified $\sum_{ii'} P_{ii'}^{\alpha\beta\gamma}$ in $S_1$ and $L_1$ and between $S_1$ and $L_1$ magnetic blocks for both hexaferrite systems.

|  | $S_1$-$S_1$ | $L_1$-$L_1$ | $S_1$-$L_1$ |
|---|---|---|---|
| Y-type | $3\begin{pmatrix} 0,0,0 & 0,0,0 & c_0,0,0 \\ 0,0,0 & 0,0,0 & 0,c_0,0 \\ -c_0,0,0 & 0,-c_0,0 & 0,0,0 \end{pmatrix}$ | $3\begin{pmatrix} 0,0,0 & 0,0,0 & c_1,0,0 \\ 0,0,0 & 0,0,0 & 0,c_1,0 \\ -c_1,0,0 & 0,-c_1,0 & 0,0,0 \end{pmatrix}$ | $3\begin{pmatrix} a_2,0,b_2 & 0,-a_2,0 & c_2,0,0 \\ 0,-a_2,0 & -a_2,0,b_2 & 0,c_2,0 \\ c_3,0,0 & 0,c_3,0 & 0,0,d_2 \end{pmatrix}$ |
| Z-type | $3\begin{pmatrix} 0,0,0 & 0,0,0 & c_0,0,0 \\ 0,0,0 & 0,0,0 & 0,c_0,0 \\ -c_0,0,0 & 0,-c_0,0 & 0,0,0 \end{pmatrix}$ | $3\begin{pmatrix} a_1,0,0 & 0,-a_1,0 & c_1,0,0 \\ 0,-a_1,0 & -a_1,0,0 & 0,c_1,0 \\ -c_1,0,0 & 0,-c_1,0 & 0,0,0 \end{pmatrix}$ | $3\begin{pmatrix} a_2,0,b_2 & 0,-a_2,0 & c_2,0,0 \\ 0,-a_2,0 & -a_2,0,b_2 & 0,c_2,0 \\ c_3,0,0 & 0,c_3,0 & 0,0,d_2 \end{pmatrix}$ |

Next, by substituting Eq. (7), Tables 1 and 2 into Eq. (2) and summing over one magnetic unit cell of TC phases, we calculated the angle $\phi$ dependent total polarization $P_{tot}(\phi)$. Contribution of each term to the polarization can be obtained separately, as shown in Tables 3 and 4.

Table 3. The calculated net polarization in each kind of magnetic block according to the simplified $\sum_i P_{ii}^{\alpha\beta\gamma}$ in Table 1 for both hexaferrite systems.

|        | S | L |
|--------|---|---|
| Y-type | 0 | 0 |
| Z-type | 0 | $\frac{2}{3}a_0\mu_L^x\mu_L^y(-\sin 2\phi, \cos 2\phi, 0)$ |

Table 4. The calculated net polarization from the inter and intra-blocks according to the simplified $\sum_{ii'} P_{ii'}^{\alpha\beta\gamma}$ in Table 2 for both hexaferrite systems.

|        | S-S | L-L | S-L |
|--------|-----|-----|-----|
| Y-type | 0   | 0   | $\frac{1}{3}c_3\mu_S^z\mu_L^y(\sin\phi, \cos\phi, 0)$ |
| Z-type | 0   | $\frac{1}{3}a_1\mu_L^x\mu_L^y(-\sin 2\phi, \cos 2\phi, 0)$ | $-\frac{1}{6}a_2\mu_S^x\mu_L^y(-\sin 2\phi, \cos 2\phi, 0)$ $+\frac{1}{3}c_3\mu_S^Z\mu_L^y(\sin\phi, \cos\phi, 0)$ |

For the Y-type hexaferrite with TC phase, the angle dependent polarization $P_Y(\phi)$ only comes from inter-block two-spin term:

$$P_Y(\phi) = P_{SL}(\phi) = \frac{1}{3}c_3\mu_S^z\mu_L^y(\sin\phi, \cos\phi, 0) \qquad (8)$$

where $c_3$ has the form $c_3 = \frac{P_{ii'}^{zxx} + P_{ii'}^{zyy}}{2}$, $i$ and $i'$ is a spin-pair between adjacent $L$ and $S$ blocks (see Supplementary Note 2). Equation (8) predicts that the $P$ vector rotates coordinately under in-plane $H$-rotation with $P\perp H\perp[001]$ relationship, fully consistent with the experimental observations in Figs. 2c&d for BSZFAO as well as many other Y-type ME hexaferrite systems. Note that, non-zero $c_3$ only allows in the spin-current mechanism (see Eq. (3)), not in the exchange-striction mechanism (see Eq. (4)), ruling out the possible cooperative contribution of the magnetostriction to the polarization of Y-type[18].

For TC phase induced polarization $P_Z$ in the Z-type hexaferrite, things are quite different: 1) The single-spin term in L blocks contributes nonzero polarization $P_L$ with sinusoidal $2\phi$ dependent:

$$P_L(\phi) = \frac{2}{3} a_0 \mu_L^x \mu_L^y (-\sin 2\phi, \cos 2\phi, 0) \tag{9}$$

where $a_0$ has the form $a_0 = \frac{P_{ii}^{xxx} - P_{ii}^{xyy} - P_{ii}^{yxy} - P_{ii}^{yyx}}{4}$, $i$ is a site in L block. 2) The two-spin terms also generate non-zero polarization between L-L ($P_{LL}$) with sinusoidal $2\phi$ dependence and L-S blocks ($P_{LS}$) with both sinusoidal $\phi$ and $2\phi$ dependences:

$$P_{LL}(\phi) = \frac{1}{3} a_1 \mu_L^x \mu_L^y (-\sin 2\phi, \cos 2\phi, 0),$$

$$P_{LS}(\phi) = -\frac{1}{6} a_2 \mu_S^x \mu_L^y (-\sin 2\phi, \cos 2\phi, 0) + \frac{1}{3} c_3 \mu_S^z \mu_L^y (\sin \phi, \cos \phi, 0) \tag{10}$$

where $a_1$, $a_2$ and $c_3$ have the form $a_1 = \frac{P_{ij}^{xxx} - P_{ij}^{xyy} - P_{ij}^{yxy} - P_{ij}^{yyx}}{4}$,

$a_2 = \frac{P_{ii'}^{xxx} - P_{ii'}^{xyy} - P_{ii'}^{yxy} - P_{ii'}^{yyx}}{4}$ and $c_3 = \frac{P_{ii'}^{zxx} + P_{ii'}^{zyy}}{2}$, respectively, $ij$ is a spin-pair between two layer of L block and $ii'$ is a spin-pair between two S and L blocks. Then, the $\phi$ dependent total polarization $P_Z$ is:

$$P_Z(\phi) = \frac{1}{6}(4 a_0 \mu_L^x + 2 a_1 \mu_L^x - a_2 \mu_S^x) \mu_L^y (-\sin 2\phi, \cos 2\phi, 0) + \frac{1}{3} c_3 \mu_S^z \mu_L^y (\sin \phi, \cos \phi, 0) \tag{11}$$

This formula predicts that if $H$ rotates clockwise in the plane with an angular speed of ω, both $P_x$ and $P_y$ in TC phase will have sinusoidal ϕ and 2ϕ dependent behaviors together, leading to a clockwise rotating of P component with a speed of ω and a counter clockwise rotating of P component with the double speed of 2ω. However, from our experiment results in Figs. 2f and 3d-g, both $P_x$ and $P_y$ shows dominant sinusoidal 2ϕ behavior and the dominant P vector rotates counterclockwise with nearly double rotating speed (Fig. 3h), indicating a relatively small sinusoidal ϕ component. This fact is also revealed by the weak asymmetric P(H) profile in Figs. 2e and 3a-b where the effect of $H_x$ reversal can be regarded as ϕ=0→ϕ=π in Eq. 11. Indeed, the weak asymmetric P(H) is a universal feature in our Z-type BSCFO and many other Z-type hexaferrite systems. Then, we will try to understand the microscopic origin of each P contribution.

It should be mentioned here that this kind of comparison between model calculation and *H* rotating experiment may be applicable to other multiferroic hexaferrites like *M*-type and *U*-type hexaferrites to check any possibility of non-spin-current mechanism since they all have structure **R** block in their lattice. Especially, under certain circumstances, they have shown irreversible polarization under *H* reversal[25,30].

## Discussion

To have the single-spin mechanism contributed $P_L$, $a_0 = \dfrac{P_{ii}^{xxx} - P_{ii}^{xyy} - P_{ii}^{yxy} - P_{ii}^{yyx}}{4}$ at some low symmetric Fe/Co sites in *L* blocks should be non-zero. There are 10 different atom positions for Fe/Co ions, which are labeled as Me1 to Me10 (see Table 5 and Fig. 1b). We only have to consider sites Me3-10 which belong to the *L* block. To deduce the non-zero components of $P_{ii}^{\alpha\beta\gamma}$ for each site, one has to consider their global site symmetry of each Wyckoff position instead of their local environment. As shown in Table 5, there are three kinds of local environment, octahedral, tetrahedron and bipyramid which allows very high local symmetries. If MeO polyhedrons are far away from each other, then those local symmetries would be a good enough approximation to calculate the $P_{ii}^{\alpha\beta\gamma}$ and net $P_L$ from each site. This is exemplified in the case of $Ba_2CoGe_2O_7$ where the $CoO_4$ tetrahedrons are separated by $Ba^{2+}$ ions so that the net *P* within a $CoO_4$ due to *p-d* hybridization mechanism can be reliably calculated without considering global symmetry[13]. However, it is not the case in the Z-type hexaferrites owing to the compact edge or corner sharing between those polyhedrons. From the site symmetry shown in Table 5, all the Me3 to Me10 sites seem to be able to have non-zero $a_0$ (see Supplementary Note 3) unless there are other hidden or accidental symmetry constraints. Therefore, at least each Fe/Co site in the *L* block can generate a non-zero net *P* via single spin mechanism, i.e., *p-d* hybridization mechanism. However, due to the quenched orbital moments in the tetrahedral and centered octahedral[31], the *p-d* hybridization in Me5-8 and Me10 would be very weak, leading to the negligible polarization by this mechanism. But, the off-center Fe/Co in octahedral can induce large orbital moments[31] which may enhance the extent of *p-d* hybridization. Therefore, Me3, Me4, and Me9 sites may provide larger $a_0$ and subsequently larger *P* via the *p-d* hybridization mechanism. This is consistent with the observations from *in-situ* X-ray diffraction[23]. However, we do not exclude the contributions to $P_L$ from other exotic single-spin mechanisms.

Table 5. The ten independent Wyckoff positions of the Fe/Co ions and their global and local symmetries.

| Atom position | Wyckoff letter | Site symmetry | Local environment |
|---|---|---|---|
| Me1 | 2a | -3m | octahedral |
| Me2 | 4f | 3m | tetrahedral |
| Me3 | 4e | 3m | octahedral |
| Me4 | 12k | m | octahedral |
| Me5 | 4e | 3m | tetrahedral |
| Me6 | 4f | 3m | octahedral |
| Me7 | 4f | 3m | tetrahedral |
| Me8 | 12k | m | octahedral |
| Me9 | 4f | 3m | octahedral |
| Me10 | 2c | -6m2 | bipyramid |

To have the two-spin mechanism contributed $P_{LL}$ and $P_{LS}$, $a_1 = \frac{P_{ij}^{xxx} - P_{ij}^{xyy} - P_{ij}^{yxy} - P_{ij}^{yyx}}{4}$, $a_2 = \frac{P_{ii'}^{xxx} - P_{ii'}^{xyy} - P_{ii'}^{yxy} - P_{ii'}^{yyx}}{4}$ and $c_3 = \frac{P_{ii'}^{zxx} + P_{ii'}^{zyy}}{2}$ should be non-zero, where *ij* is a spin-pair between two layer of *L* block and *ii'* is a spin-pair between two *S* and *L* blocks. As discussed in the case of Y-type, only the sinusoidal ϕ component and $c_3$ corresponds to the spin-current generated polarization. In the case of Z-type, the sinusoidal ϕ component and $c_3$ is very weak which may be due to the much lower volume density of the *L-S* block boundaries in the Z-type than that of Y-type. As for the $a_1$ and $a_2$, they are exactly zero for both the exchange-striction and spin-current mechanisms (see Eqs. (3) and (4)). That means, other exotic mechanisms must be the origins of sin2ϕ/cos2ϕ components in $P_{LL}$ and $P_{LS}$ if they are non-zero in reality. Actually, a recently proposed anisotropic symmetric exchange mechanism due to spin orbital coupling allows non-zero $a_1$ and $a_2$[32]. To examine this mechanism, a first principle calculation on BSCFO is highly desired.

In conclusion, we compared the responses of polarization to the external magnetic field between Y-type and Z-type hexaferrites, and revealed a direct link between the symmetry in magnetic $L$ blocks and the microscopic mechanisms of transverse cone induced polarization and related ME behaviors. To prove this, we only rely on a rigorous lattice symmetry-based local ME tensors approach without requiring any first principal calculation. This means the lattice symmetry, especially local symmetry, is also crucial in determine the microscopic mechanism of polarization and related ME behavior in a spin-driven multiferroics. More important, we have suggested the unexpected contribution of polarization from *p-d* hybridization and other exotic mechanisms in Z-type hexaferrites which is beyond any previous expectations for multiferroic with conical spin-order.

**ACKNOWLEDGMENTS**

We thank H. J. Xiang and J. H. Han for enlightening discussions. This work was supported by the National Creative Research Initiative (2010-0018300) through the National Research Foundation (NRF) funded by the Korea government. The work at China was supported by the National Natural Science Foundation of China under Grant Nos. 11374347, 11674384.

## Methods

**Sample preparation and x-ray diffraction**. Single crystals of Y-type hexaferrite with a nominal composition of $Ba_{0.5}Sr_{1.5}Zn_2(Fe_{0.92}Al_{0.08})_{12}O_{22}$ and Z-type hexaferrite with a nominal composition of $Ba_{0.52}Sr_{2.48}Co_2Fe_{24}O_{41}$ were grown from the $Na_2O$-$Fe_2O_3$ flux in the air. The crystals were collected by checking the *c*-axis lattice parameter from the X-ray diffraction study. The orientation of the single crystals was determined using back-reflection X-ray Laue photographs. Before any magnetic and electrical measurements, all the samples were heat-treated to remove oxygen vacancy at 900 °C under flowing $O_2$ for 8 days.

**Magnetic and electric measurements**. The sample was cut into a rectangular parallelepiped shape with the large surfaces normal to the [100], [120] or both directions. Each face was mechanically polished to obtain a flat smooth surface. Electrodes were formed on two faces normal to [100] or four faces perpendicular to [100] and [120] directions. To measure the polarization, each specimen was subjected to an ME annealing procedure starting at 120 K for Y-type and 305 K for Z-type hexaferrites. Here, we introduce a Cartesian coordinate as shown in Fig. 2a; the *x*, *y*, and *z* axes are parallel to the [100], [120] and [001] directions, respectively. For the ME annealing condition of the $\pm E_y H_x$, the sample was electrically poled at about $E_y = \pm 120$ kVm$^{-1}$ upon changing the magnetic field from the paraelectric collinear state ($H_x$=20 kOe) to the ferrimagnetic state ($H_x$ = 10 kOe). Then, with the same $E_y$ and the $H_x$, the sample was cooled to the targeted temperatures (30 K for Y-type and 305 & 10 K for Z-type). *P* were measured by integrating the displacement currents generated between one or two pairs of electrodes while sweeping *H* or rotating sample under constant *H* using piezoelectric rotator.

**References**

[1] Kimura,T., Goto,T., Shintani, H., Ishizaka, K., Arima, T. and Tokura, Y. Magnetic control of ferroelectric polarization. *Nature* **426**, 55 (2003).

[2] Wang, J. et al. Epitaxial $BiFeO_3$ multiferroic thin film heterostructures. *Science* **299**, 1719 (2003).

[3] Chapon, L. C. et al. Structural anomalies and multiferroic behavior in magnetically frustrated $TbMn_2O_5$. *Phys. Rev. Lett*. **93**, 177402 (2004).

[4] Fiebig, M. Revival of the magnetoelectric effect. *J. Phys. D* **38**, R123–R152 (2005).

[5] Spaldin, N. A. and Fiebig, M. The renaissance of magnetoelectric multiferroics. *Science* **309**, 391–392 (2005).

[6] Tokura, Seki, Y. S. and Nagaosa, N. Multiferroics of spin origin. *Rep. Prog. Phys.* **77**, 076501 (2014).

[7] Katsura, H., Nagaosa, N. and Balatsky, A. V. Spin current and magnetoelectric effect in noncollinear magnets. *Phys. Rev. Lett.* **95**, 057205 (2005).

[8] Mostovoy, M. Ferroelectricity in spiral magnets. *Phys. Rev. Lett.* **96**, 067601 (2006).

[9] Greenwald, S. and Smart, J. S. Deformations in the crystal structures of anti-ferromagnetic compounds. *Nature* **166**, 523 (1950).

[10] Smart J. S. and Greenwald, S. Crystal structure transitions in antiferromagnetic compounds at the Curie temperature. *Phys. Rev.* **82**, 113 (1951).

[11] Jia, C., Onoda, S., Nagaosa, N. and Han, J. H. Bond electronic polarization induced by spin. *Phys. Rev. B* **74**, 224444 (2006).

[12] Arima, T. J. Ferroelectricity induced by proper-screw type magnetic order. *J. Phys. Soc. Japan* **76**, 73702 (2007).

[13] Murakawa, H. et al. Ferroelectricity induced by spin-dependent metal-ligand hybridization in $Ba_2CoGe_2O_7$. *Phys. Rev. Lett.* **105** 137202 (2010).

[14] Newnham, R. E. Properties of Materials, *Oxford University Press*, New York, NY, USA (2005).

[15] Kimura, T. Magnetoelectric hexaferrites. *Annu. Rev. Condens. Matter Phys.* **3**,

93-110 (2012).

[16] Kimura, T., Lawes, G. and Ramirez, A. Electric polarization rotation in a hexaferrite with long-wavelength magnetic structures. *Phys. Rev. Lett*. **94**, 137201 (2005).

[17] Kitagawa, Y. et al. Low-field magnetoelectric effect at room temperature. *Nature Mater.* **9**, 797–802 (2010).

[18] Soda, M., Ishikura, T., Nakamura, H., Wakabayashi, Y. and Kimura, T. Magnetic ordering in relation to the room-temperature magnetoelectric effect of $Sr_3Co_2Fe_{24}O_{41}$. *Phys. Rev. Lett.* **106**, 087201 (2011).

[19] Momozawa, N., Yamaguchi, Y. and Mita, M. Magnetic structure change in $Ba_2Mg_2Fe_{12}O_{22}$. *J. Phys. Soc. Jpn.* **55**, 1350 (1986).

[20] Chun, S. H. et al. Realization of giant magnetoelectricity in helimagnets. *Phys. Rev. Lett.* **104**, 037204 (2010).

[21] Yamasaki, Y. et al. Magnetic reversal of the ferroelectric polarization in a multiferroic spinel oxide. *Phys. Rev. Lett.* **96**, 207204 (2006).

[22] Ishiwata, S., Taguchi, Y., Murakawa, H., Onose, Y. and Tokura, Y. Low-magnetic-field control of electric polarization vector in a helimagnet. *Science* **319**, 1643 (2008).

[23] Chun, S. H. et al. Microscopic observation of entangled multi-magnetoelectric coupling phenomenon. arXiv:1706.01144v1

[24] Chun, S. H. et al. Electric field control of nonvolatile four-state magnetization at room temperature. *Phys. Rev. Lett.* **108**, 177201 (2012).

[25] Tokunaga, Y. et al. Multiferroic M-type hexaferrites with a room-temperature conical state and magnetically controllable spin helicity. *Phys. Rev. Lett*. **105**, 257201 (2010).

[26] Kurumaji, T. et al. Magnetic-field induced competition of two multiferroic orders in a triangular-lattice helimagnet $MnI_2$. *Phys. Rev. Lett.* **106**, 167206 (2011).

[27] Xiang, H. J., Kan, E. J., Zhang, Y., Whangbo, M.-H. and Gong, X. G. General theory for the ferroelectric polarization induced by spin-spiral order. *Phys. Rev. Lett.* **107**, 157202 (2011).

[28] Kaplan, T. A. and Mahanti, S. D. Canted-spin-caused electric dipoles: A local symmetry theory. *Phys. Rev. B* **83**, 174432 (2011).

[29] Miyahara, S. and Furukawa, N. Theory of antisymmetric spin-pair-dependent electric polarization in multiferroics. *Phys. Rev. B* **93**, 014445 (2016).

[30] Okumura, K. et al. Magnetism and magnetoelectricity of a U-type hexaferrite $Sr_4Co_2Fe_{36}O_{60}$. *Appl. Phys. Lett.* **98**, 212504 (2011).

[31] Noh, W.-S. et al. Magnetic origin of giant magnetoelectricity in doped Y-type hexaferrite $Ba_{0.5}Sr_{1.5}Zn_2(Fe_{1-x}Al_x)_{12}O_{22}$. *Phys. Rev. Lett.* **114**, 117603 (2015).

[32] Feng, J. S. and Xiang, H. J. Anisotropic symmetric exchange as a new mechanism for multiferroicity. *Phys. Rev. B* **93**, 174416 (2016).

Figure Captions

**Fig. 1** The crystal and magnetic structure of **a** Y-type hexaferrite and **b** Z-type hexaferrite. The Schematic models of the transverse cone are shown for both systems. In the left panel of **b**, 10 different Wyckoff positions for Fe/Co ions are labeled as Me1 to Me10.

**Fig. 2** Schematics showing relationship between the direction of $P$, $H$, $k_0$, the crystallographic axes, and the Cartesian coordinate under **a** $H$ reversal and **b** $H$ rotation, respectively. $P_y$ curves of Y-type BSZAFO measured under **c** $H_x$ sweeping and **d** $H$ rotation ($H$ = 0.2 kOe and 2 kOe) at 30 K. $\Delta P_y$ curves of Z-type BSCFO measured under **e** $H_x$ sweeping and **f** $H$ rotation ($H$ = 5 kOe and 12 kOe) at 305 K.

**Fig. 3** $H$ dependent $P$ vector measurements at 10 K. **a** $P_x$ and **b** $P_y$ curves of Z-type BSCFO measured under $H_x$ sweeping. **c** Schematics of $H$ rotating measurement configuration. Angle dependent $P_x$ and $P_y$ curves measured under rotating $H$ of (**d**, **e**) 4 kOe and (**f**, **g**) 10 kOe. **h** Schematic angle dependent relationship between $H$ and $P$ vectors.

**Fig. 4** Crystal structure models of the Y-type and Z-type. **a** Simplified generic models for the crystal structures of the Y-type and Z-type on the basis of the magnetic block approximation. The trajectory of the transverse cone spin configuration in $yz$ plane for $H//x$ condition is shown in the right panel. The schematic illustration of the symmetry operations for **b** one of the layers in the S block as a representative, **c** two layers in the different blocks or in the same block.

Fig. 1 Y. S. Chai et al.

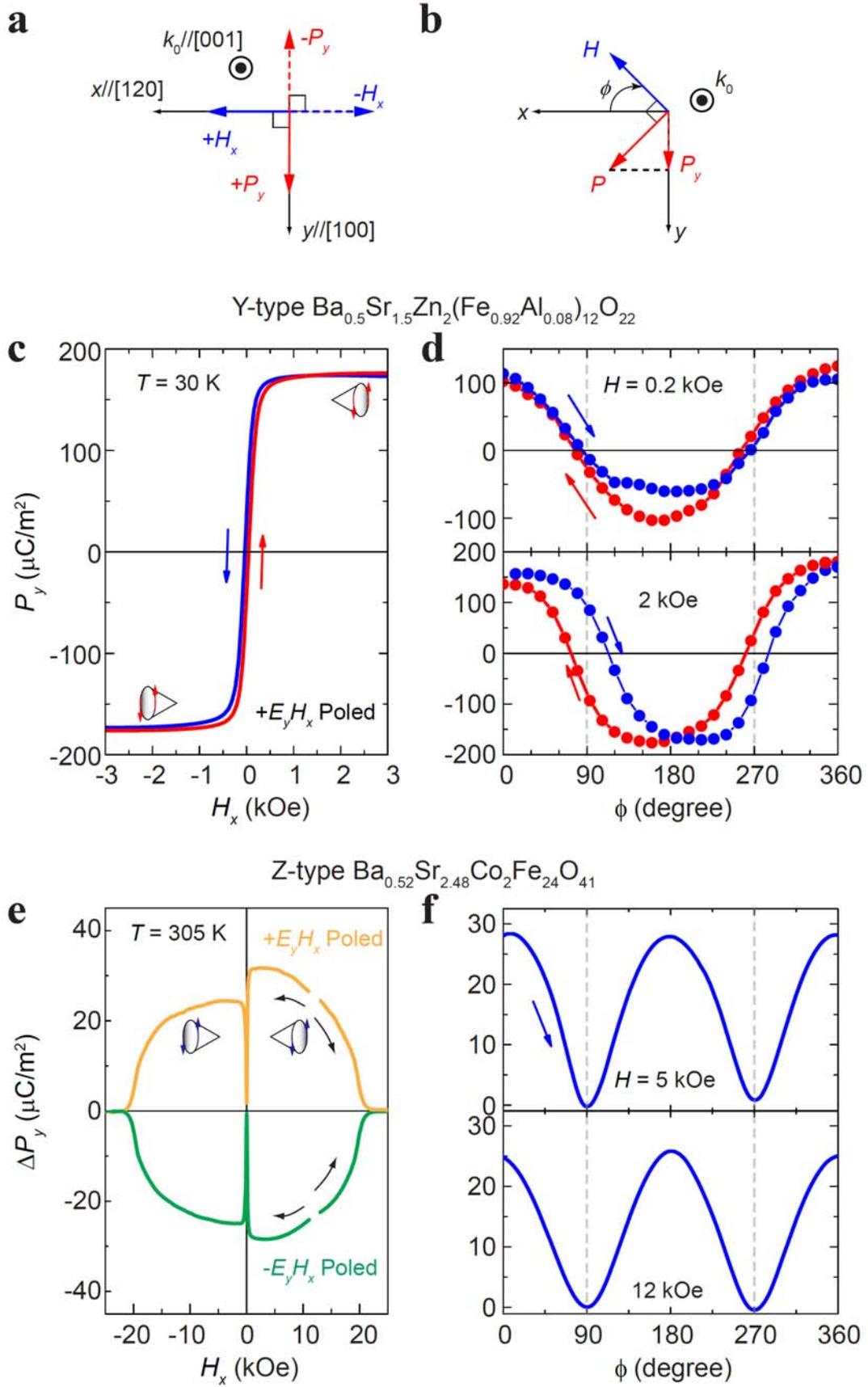

Fig. 2 Y. S. Chai et al.

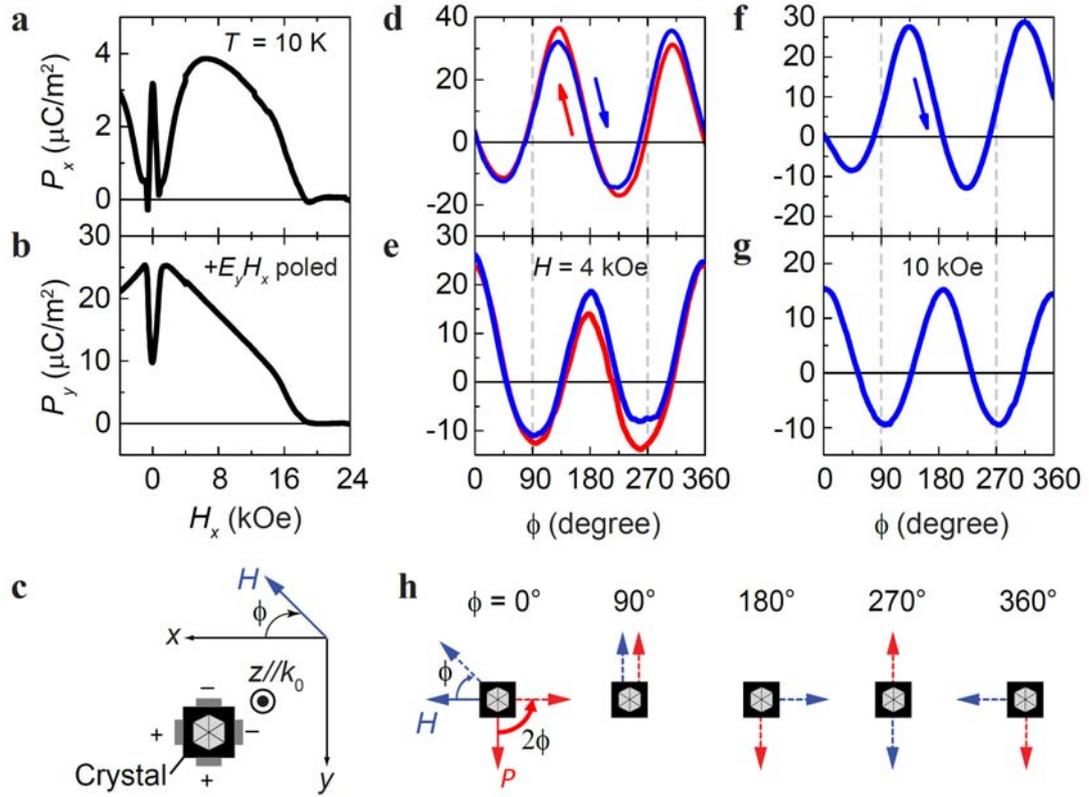

Fig. 3 Y. S. Chai et al.

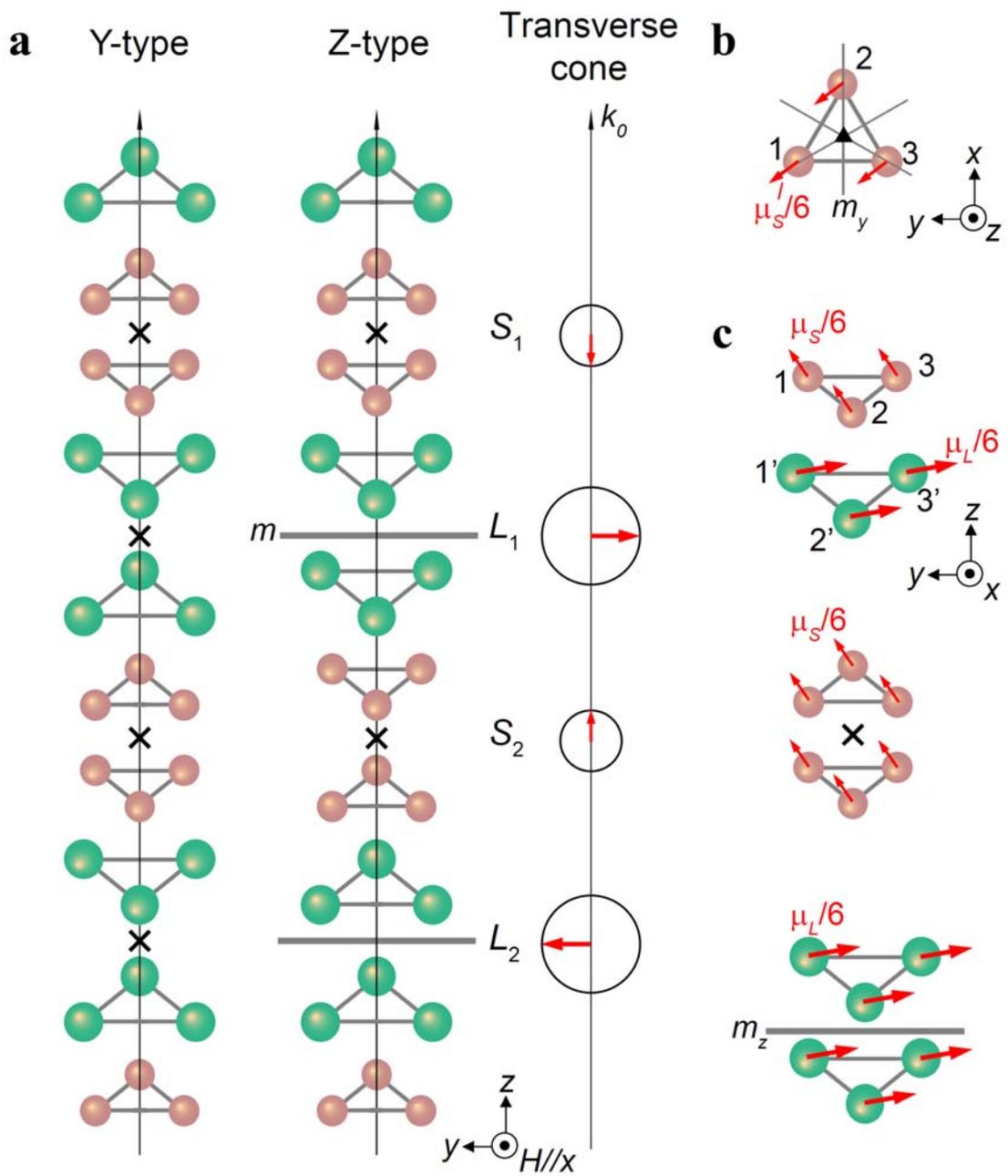

Fig. 4 Y. S. Chai et al.

## Supplementary Note 1: PROPERTIES OF THE ME LOCAL TENSORS

According to the definition of $P_{ii}^{\alpha\beta\gamma}$ and $P_{ij}^{\alpha\beta\gamma}$ in Eq. (1), they are the even function of time because electric polarization is an even function of time. Therefore, both ME tensor are third rank **polar** tensors with the transformation laws:

$$P_{pp}^{ijk}{'} = a_{il}a_{jm}a_{kn}P_{pp}^{lmn} \quad \text{and} \quad P_{pq}^{ijk}{'} = a_{il}a_{jm}a_{kn}P_{pq}^{lmn} \tag{1}$$

where $a_{il}$, $a_{jm}$ and $a_{kn}$ are the direction cosines relating the two coordinate systems. From Supplementary Eq. (1), we could calculate the transformed matrix form of the $P_{ii}^{\alpha\beta\gamma}$ and $P_{ij}^{\alpha\beta\gamma}$ under a symmetry operation. In Supplementary Table 1, we list the transformed matrix forms of $P_{ij}^{\alpha\beta\gamma}$ under the selected symmetry operations required for all the 32 crystal classes. The transformation of $P_{ii}^{\alpha\beta\gamma}$ under any symmetry operation can be deduced since $P_{ii}^{\alpha\beta\gamma} = P_{ii}^{\beta\alpha\gamma}$. These are the transformation operations needed to develop the local ME matrices for various multiferroic systems if people want to perform similar calculations in systems other than hexaferrites. From Supplementary Table 1, the matrices form of $P_{ii}^{\alpha\beta\gamma}$ and $P_{ij}^{\alpha\beta\gamma}$ can be simplified by applying these symmetry operations according to Neumann's Principle:

$$P_{pp}^{ijk}{'} = P_{pp}^{ijk} \quad \text{and} \quad P_{pq}^{ijk}{'} = P_{pq}^{ijk} \tag{2}$$

if the single spin or the spin-pair possesses the symmetry operation used in the transformation.

Supplementary Table 1. The transformed matrices of a two-spin tensor $P_{12}^{\alpha\beta\gamma}$ under various symmetry operations.

|  | Transformed $P_{12}^{\alpha\beta\gamma}$ in the matrix form |
|---|---|
| **1** | $\begin{pmatrix} P_{12}^{xx(x)}, P_{12}^{xx(y)}, P_{12}^{xx(z)} & P_{12}^{xy(x)}, P_{12}^{xy(y)}, P_{12}^{xy(z)} & P_{12}^{xz(x)}, P_{12}^{xz(y)}, P_{12}^{xz(z)} \\ P_{12}^{yx(x)}, P_{12}^{yx(y)}, P_{12}^{yx(z)} & P_{12}^{yy(x)}, P_{12}^{yy(y)}, P_{12}^{yy(z)} & P_{12}^{yz(x)}, P_{12}^{yz(y)}, P_{12}^{yz(z)} \\ P_{12}^{zx(x)}, P_{12}^{zx(y)}, P_{12}^{zx(z)} & P_{12}^{zy(x)}, P_{12}^{zy(y)}, P_{12}^{zy(z)} & P_{12}^{zz(x)}, P_{12}^{zz(y)}, P_{12}^{zz(z)} \end{pmatrix}$ |
| **-1** | $\begin{pmatrix} -P_{12}^{xx(x)} -P_{12}^{xx(y)} -P_{12}^{xx(z)} & -P_{12}^{xy(x)} -P_{12}^{xy(y)} -P_{12}^{xy(z)} & -P_{12}^{xz(x)} -P_{12}^{xz(y)} -P_{12}^{xz(z)} \\ -P_{12}^{yx(x)} -P_{12}^{yx(y)} -P_{12}^{yx(z)} & -P_{12}^{yy(x)} -P_{12}^{yy(y)} -P_{12}^{yy(z)} & -P_{12}^{yz(x)} -P_{12}^{yz(y)} -P_{12}^{yz(z)} \\ -P_{12}^{zx(x)} -P_{12}^{zx(y)} -P_{12}^{zx(z)} & -P_{12}^{zy(x)} -P_{12}^{zy(y)} -P_{12}^{zy(z)} & -P_{12}^{zz(x)} -P_{12}^{zz(y)} -P_{12}^{zz(z)} \end{pmatrix}$ |
| **2//x** ($2_x$) | $\begin{pmatrix} P_{12}^{xx(x)} -P_{12}^{xx(y)} -P_{12}^{xx(z)} & -P_{12}^{xy(x)} P_{12}^{xy(y)} P_{12}^{xy(z)} & -P_{12}^{xz(x)} P_{12}^{xz(y)} P_{12}^{xz(z)} \\ -P_{12}^{yx(x)} P_{12}^{yx(y)} P_{12}^{yx(z)} & P_{12}^{yy(x)} -P_{12}^{yy(y)} -P_{12}^{yy(z)} & P_{12}^{yz(x)} -P_{12}^{yz(y)} -P_{12}^{yz(z)} \\ -P_{12}^{zx(x)} P_{12}^{zx(y)} P_{12}^{zx(z)} & P_{12}^{zy(x)} -P_{12}^{zy(y)} -P_{12}^{zy(z)} & P_{12}^{zz(x)} -P_{12}^{zz(y)} -P_{12}^{zz(z)} \end{pmatrix}$ |
| **2//y** ($2_y$) | $\begin{pmatrix} -P_{12}^{xx(x)} P_{12}^{xx(y)} -P_{12}^{xx(z)} & P_{12}^{xy(x)} -P_{12}^{xy(y)} P_{12}^{xy(z)} & -P_{12}^{xz(x)} P_{12}^{xz(y)} -P_{12}^{xz(z)} \\ P_{12}^{yx(x)} -P_{12}^{yx(y)} P_{12}^{yx(z)} & -P_{12}^{yy(x)} P_{12}^{yy(y)} -P_{12}^{yy(z)} & P_{12}^{yz(x)} -P_{12}^{yz(y)} P_{12}^{yz(z)} \\ -P_{12}^{zx(x)} P_{12}^{zx(y)} -P_{12}^{zx(z)} & P_{12}^{zy(x)} -P_{12}^{zy(y)} P_{12}^{zy(z)} & -P_{12}^{zz(x)} P_{12}^{zz(y)} -P_{12}^{zz(z)} \end{pmatrix}$ |
| **2//z** ($2_z$) | $\begin{pmatrix} -P_{12}^{xx(x)} -P_{12}^{xx(y)} P_{12}^{xx(z)} & -P_{12}^{xy(x)} -P_{12}^{xy(y)} P_{12}^{xy(z)} & P_{12}^{xz(x)} P_{12}^{xz(y)} -P_{12}^{xz(z)} \\ -P_{12}^{yx(x)} -P_{12}^{yx(y)} P_{12}^{yx(z)} & -P_{12}^{yy(x)} -P_{12}^{yy(y)} P_{12}^{yy(z)} & P_{12}^{yz(x)} P_{12}^{yz(y)} -P_{12}^{yz(z)} \\ P_{12}^{zx(x)} P_{12}^{zx(y)} -P_{12}^{zx(z)} & P_{12}^{zy(x)} P_{12}^{zy(y)} -P_{12}^{zy(z)} & -P_{12}^{zz(x)} -P_{12}^{zz(y)} P_{12}^{zz(z)} \end{pmatrix}$ |
| **m⊥x** ($m_x$) | $\begin{pmatrix} -P_{12}^{xx(x)} P_{12}^{xx(y)} P_{12}^{xx(z)} & P_{12}^{xy(x)} -P_{12}^{xy(y)} -P_{12}^{xy(z)} & P_{12}^{xz(x)} -P_{12}^{xz(y)} -P_{12}^{xz(z)} \\ P_{12}^{yx(x)} -P_{12}^{yx(y)} -P_{12}^{yx(z)} & -P_{12}^{yy(x)} P_{12}^{yy(y)} P_{12}^{yy(z)} & -P_{12}^{yz(x)} P_{12}^{yz(y)} P_{12}^{yz(z)} \\ P_{12}^{zx(x)} -P_{12}^{zx(y)} -P_{12}^{zx(z)} & -P_{12}^{zy(x)} P_{12}^{zy(y)} P_{12}^{zy(z)} & -P_{12}^{zz(x)} P_{12}^{zz(y)} P_{12}^{zz(z)} \end{pmatrix}$ |
| **m⊥y** ($m_y$) | $\begin{pmatrix} P_{12}^{xx(x)} -P_{12}^{xx(y)} P_{12}^{xx(z)} & -P_{12}^{xy(x)} P_{12}^{xy(y)} -P_{12}^{xy(z)} & P_{12}^{xz(x)} -P_{12}^{xz(y)} P_{12}^{xz(z)} \\ -P_{12}^{yx(x)} P_{12}^{yx(y)} -P_{12}^{yx(z)} & P_{12}^{yy(x)} -P_{12}^{yy(y)} P_{12}^{yy(z)} & -P_{12}^{yz(x)} P_{12}^{yz(y)} -P_{12}^{yz(z)} \\ P_{12}^{zx(x)} -P_{12}^{zx(y)} P_{12}^{zx(z)} & -P_{12}^{zy(x)} P_{12}^{zy(y)} -P_{12}^{zy(z)} & P_{12}^{zz(x)} -P_{12}^{zz(y)} P_{12}^{zz(z)} \end{pmatrix}$ |
| **m⊥z** ($m_z$) | $\begin{pmatrix} P_{12}^{xx(x)} P_{12}^{xx(y)} -P_{12}^{xx(z)} & P_{12}^{xy(x)} P_{12}^{xy(y)} -P_{12}^{xy(z)} & -P_{12}^{xz(x)} -P_{12}^{xz(y)} P_{12}^{xz(z)} \\ P_{12}^{yx(x)} P_{12}^{yx(y)} -P_{12}^{yx(z)} & P_{12}^{yy(x)} P_{12}^{yy(y)} -P_{12}^{yy(z)} & -P_{12}^{yz(x)} -P_{12}^{yz(y)} P_{12}^{yz(z)} \\ -P_{12}^{zx(x)} -P_{12}^{zx(y)} P_{12}^{zx(z)} & -P_{12}^{zy(x)} -P_{12}^{zy(y)} P_{12}^{zy(z)} & P_{12}^{zz(x)} P_{12}^{zz(y)} -P_{12}^{zz(z)} \end{pmatrix}$ |
| **3//[111]** | $\begin{pmatrix} P_{12}^{zz(z)}, P_{12}^{zz(x)}, P_{12}^{zz(y)} & P_{12}^{zx(z)}, P_{12}^{zx(x)}, P_{12}^{zx(y)} & P_{12}^{zy(z)}, P_{12}^{zy(x)}, P_{12}^{zy(y)} \\ P_{12}^{xz(z)}, P_{12}^{xz(x)}, P_{12}^{xz(y)} & P_{12}^{xx(z)}, P_{12}^{xx(x)}, P_{12}^{xx(y)} & P_{12}^{xy(z)}, P_{12}^{xy(x)}, P_{12}^{xy(y)} \\ P_{12}^{yz(z)}, P_{12}^{yz(x)}, P_{12}^{yz(y)} & P_{12}^{yx(z)}, P_{12}^{yx(x)}, P_{12}^{yx(y)} & P_{12}^{yy(z)}, P_{12}^{yy(x)}, P_{12}^{yy(y)} \end{pmatrix}$ |
| **4//z** | $\begin{pmatrix} -P_{12}^{yy(y)}, P_{12}^{yy(x)}, P_{12}^{yy(z)} & P_{12}^{yx(y)}, -P_{12}^{yx(x)}, -P_{12}^{yx(z)} & P_{12}^{yz(y)}, -P_{12}^{yz(x)}, -P_{12}^{yz(z)} \\ P_{12}^{xy(y)}, -P_{12}^{xy(x)}, -P_{12}^{xy(z)} & -P_{12}^{xx(y)}, P_{12}^{xx(x)}, P_{12}^{xx(z)} & -P_{12}^{xz(y)}, P_{12}^{xz(x)}, P_{12}^{xz(z)} \\ P_{12}^{zy(y)}, -P_{12}^{zy(x)}, -P_{12}^{zy(z)} & -P_{12}^{zx(y)}, P_{12}^{zx(x)}, P_{12}^{zx(z)} & -P_{12}^{zz(y)}, P_{12}^{zz(x)}, P_{12}^{zz(z)} \end{pmatrix}$ |

| -4//z | $\begin{pmatrix} P_{12}^{yy(y)}, -P_{12}^{yy(x)}, -P_{12}^{yy(z)} & -P_{12}^{yx(y)}, P_{12}^{yx(x)}, P_{12}^{yx(z)} & -P_{12}^{yz(y)}, P_{12}^{yz(x)}, P_{12}^{yz(z)} \\ -P_{12}^{xy(y)}, P_{12}^{xy(x)}, P_{12}^{xy(z)} & P_{12}^{xx(y)}, -P_{12}^{xx(x)}, -P_{12}^{xx(z)} & P_{12}^{xz(y)}, -P_{12}^{xz(x)}, -P_{12}^{xz(z)} \\ -P_{12}^{zy(y)}, P_{12}^{zy(x)}, P_{12}^{zy(z)} & P_{12}^{zx(y)}, -P_{12}^{zx(x)}, -P_{12}^{zx(z)} & P_{12}^{zz(y)}, -P_{12}^{zz(x)}, -P_{12}^{zz(z)} \end{pmatrix}$ |
|---|---|
| 3//z ($3_z$) | $P_{12}^{xx(x)}{}' = \dfrac{-P_{12}^{xx(x)} + \sqrt{3}P_{12}^{xx(y)} + \sqrt{3}P_{12}^{xy(x)} - 3P_{12}^{xy(y)} + \sqrt{3}P_{12}^{yx(x)} - 3P_{12}^{yx(y)} - 3P_{12}^{yy(x)} + 3\sqrt{3}P_{12}^{yy(y)}}{8}$ $P_{12}^{xy(x)}{}' = \dfrac{-\sqrt{3}P_{12}^{xx(x)} + 3P_{12}^{xx(y)} - P_{12}^{xy(x)} + \sqrt{3}P_{12}^{xy(y)} + 3P_{12}^{yx(x)} - 3\sqrt{3}P_{12}^{yx(y)} + \sqrt{3}P_{12}^{yy(x)} - 3P_{12}^{yy(y)}}{8}$ $P_{12}^{yx(x)}{}' = \dfrac{-\sqrt{3}P_{12}^{xx(x)} + 3P_{12}^{xx(y)} + 3P_{12}^{xy(x)} - 3\sqrt{3}P_{12}^{xy(y)} - P_{12}^{yx(x)} + \sqrt{3}P_{12}^{yx(y)} + \sqrt{3}P_{12}^{yy(x)} - 3P_{12}^{yy(y)}}{8}$ $P_{12}^{yy(x)}{}' = \dfrac{-3P_{12}^{xx(x)} + 3\sqrt{3}P_{12}^{xx(y)} - \sqrt{3}P_{12}^{xy(x)} + 3P_{12}^{xy(y)} - \sqrt{3}P_{12}^{yx(x)} + 3P_{12}^{yx(y)} - P_{12}^{yy(x)} + \sqrt{3}P_{12}^{yy(y)}}{8}$ $P_{12}^{xx(y)}{}' = \dfrac{-\sqrt{3}P_{12}^{xx(x)} - P_{12}^{xx(y)} + 3P_{12}^{xy(x)} + \sqrt{3}P_{12}^{xy(y)} + 3P_{12}^{yx(x)} + \sqrt{3}P_{12}^{yx(y)} - 3\sqrt{3}P_{12}^{yy(x)} - 3P_{12}^{yy(y)}}{8}$ $P_{12}^{xy(y)}{}' = \dfrac{-3P_{12}^{xx(x)} - \sqrt{3}P_{12}^{xx(y)} - \sqrt{3}P_{12}^{xy(x)} - P_{12}^{xy(y)} + 3\sqrt{3}P_{12}^{yx(x)} + 3P_{12}^{yx(y)} + 3P_{12}^{yy(x)} + \sqrt{3}P_{12}^{yy(y)}}{8}$ $P_{12}^{yx(y)}{}' = \dfrac{-3P_{12}^{xx(x)} - \sqrt{3}P_{12}^{xx(y)} + 3\sqrt{3}P_{12}^{xy(x)} + 3P_{12}^{xy(y)} - \sqrt{3}P_{12}^{yx(x)} - P_{12}^{yx(y)} + 3P_{12}^{yy(x)} + \sqrt{3}P_{12}^{yy(y)}}{8}$ $P_{12}^{yy(y)}{}' = \dfrac{-3\sqrt{3}P_{12}^{xx(x)} - 3P_{12}^{xx(y)} - 3P_{12}^{xy(x)} - \sqrt{3}P_{12}^{xy(y)} - 3P_{12}^{yx(x)} - \sqrt{3}P_{12}^{yx(y)} - \sqrt{3}P_{12}^{yy(x)} - P_{12}^{yy(y)}}{8}$ $P_{12}^{xz(x)}{}' = \dfrac{P_{12}^{xz(x)} - \sqrt{3}P_{12}^{xz(y)} - \sqrt{3}P_{12}^{yz(x)} + 3P_{12}^{yz(y)}}{4}$ $P_{12}^{zx(x)}{}' = \dfrac{P_{12}^{zx(x)} - \sqrt{3}P_{12}^{zx(y)} - \sqrt{3}P_{12}^{zy(x)} + 3P_{12}^{zy(y)}}{4}$ $P_{12}^{xz(y)}{}' = \dfrac{\sqrt{3}P_{12}^{xz(x)} + P_{12}^{xz(y)} - 3P_{12}^{yz(x)} - \sqrt{3}P_{12}^{yz(y)}}{4}$ $P_{12}^{zx(y)}{}' = \dfrac{\sqrt{3}P_{12}^{zx(x)} + P_{12}^{zx(y)} - 3P_{12}^{zy(x)} - \sqrt{3}P_{12}^{zy(y)}}{4}$ $P_{12}^{yz(x)}{}' = \dfrac{\sqrt{3}P_{12}^{xz(x)} - 3P_{12}^{xz(y)} + P_{12}^{yz(x)} - \sqrt{3}P_{12}^{yz(y)}}{4}$ $P_{12}^{zy(x)}{}' = \dfrac{\sqrt{3}P_{12}^{zx(x)} - 3P_{12}^{zx(y)} + P_{12}^{zy(x)} - \sqrt{3}P_{12}^{zy(y)}}{4}$ $P_{12}^{yz(y)}{}' = \dfrac{3P_{12}^{xz(x)} + \sqrt{3}P_{12}^{xz(y)} + \sqrt{3}P_{12}^{yz(x)} + P_{12}^{yz(y)}}{4}$ $P_{12}^{zy(y)}{}' = \dfrac{3P_{12}^{zx(x)} + \sqrt{3}P_{12}^{zx(y)} + \sqrt{3}P_{12}^{zy(x)} + P_{12}^{zy(y)}}{4}$ $P_{12}^{xx(z)}{}' = \dfrac{P_{12}^{xx(z)} - \sqrt{3}P_{12}^{xy(z)} - \sqrt{3}P_{12}^{yx(z)} + 3P_{12}^{yy(z)}}{4}$ |

$$P_{12}^{xy(z)'} = \frac{\sqrt{3}P_{12}^{xx(z)} + P_{12}^{xy(z)} - 3P_{12}^{yx(z)} - \sqrt{3}P_{12}^{yy(z)}}{4}$$

$$P_{12}^{yx(z)'} = \frac{\sqrt{3}P_{12}^{xx(z)} - 3P_{12}^{xy(z)} + P_{12}^{yx(z)} - \sqrt{3}P_{12}^{yy(z)}}{4}$$

$$P_{12}^{yy(z)'} = \frac{3P_{12}^{xx(z)} + \sqrt{3}P_{12}^{xy(z)} + \sqrt{3}P_{12}^{yx(z)} + P_{12}^{yy(z)}}{4}$$

$$P_{12}^{xz(z)'} = \frac{-P_{12}^{xz(z)} + \sqrt{3}P_{12}^{yz(z)}}{2}, \quad P_{12}^{zx(z)'} = \frac{-P_{12}^{zx(z)} + \sqrt{3}P_{12}^{zy(z)}}{2}$$

$$P_{12}^{yz(z)'} = \frac{-\sqrt{3}P_{12}^{xz(z)} - P_{12}^{yz(z)}}{2}, \quad P_{12}^{zy(z)'} = \frac{-\sqrt{3}P_{12}^{zx(z)} - P_{12}^{zy(z)}}{2}$$

$$P_{12}^{zz(x)'} = \frac{-P_{12}^{zz(x)} + \sqrt{3}P_{12}^{zz(y)}}{2}, \quad P_{12}^{zz(y)'} = \frac{-\sqrt{3}P_{12}^{zz(x)} - P_{12}^{zz(y)}}{2}, \quad P_{12}^{zz(z)'} = P_{12}^{zz(z)}$$

## Supplementary Note 2: THE SIMPLIFICATION OF LOCAL ME TENSORS IN HEXAFERRITES

As shown in Fig. 4(b) and according to Eq. (1), the total local electric dipole from one S layer due to three identical spins $\mu_S/6$ at site 1, 2 and 3 via single spin tensor term can be expressed in Einstein convention as:

$$\sum_i p_{ii}^{\gamma} = (1/6)^2 \sum_i P_{ii}^{\alpha\beta\gamma} \mu_S^{\alpha} \mu_S^{\beta} = (1/6)^2 \mu_S^{\alpha} \mu_S^{\beta} (\sum_i P_{ii}^{\alpha\beta\gamma}) \tag{3}$$

where $i = 1, 2$ and $3$. We further noticed that this layer has a $3_z$ and my a three-fold rotation along $z$-axis and three mirrors including $z$-axis (one of them is $m_y$) are also allowed for. From Neumann's Principle (Supplementary Eq. 2), the transformed and untransformed sum of the single-spin tensor $\sum_i P_{ii}^{\alpha\beta\gamma}$ in one layer must be equal:

$$3_z(\sum_i P_{ii}^{\alpha\beta\gamma}) = (\sum_i P_{ii}^{\alpha\beta\gamma}) \text{ and } m_y(\sum_i P_{ii}^{\alpha\beta\gamma}) = (\sum_i P_{ii}^{\alpha\beta\gamma}) \tag{4}$$

Therefore, from Supplementary Table 1, we could deduce a much-simplified matrix of $\sum_i P_{ii}^{\alpha\beta\gamma}$:

$$\sum_i P_{ii}^{\alpha\beta\gamma} = 3\begin{pmatrix} a,0,b & 0,-a,0 & c,0,0 \\ 0,-a,0 & -a,0,b & 0,c,0 \\ c,0,0 & 0,c,0 & 0,0,d \end{pmatrix}, \quad (5)$$

$$a = \frac{P_{11}^{xxx} - P_{11}^{xyy} - P_{11}^{yxy} - P_{11}^{yyx}}{4}, b = \frac{P_{11}^{xxz} + P_{11}^{yyz}}{2}, c = \frac{P_{11}^{xzx} + P_{11}^{yzy}}{2}, d = P_{11}^{zzz}$$

where $P_{11}^{\alpha\beta\gamma}$ are the matrix components of single-spin tensor at site 1.

On the other hand, the total local dipoles by three inter-block spin-pairs 11', 22' and 33' via two-spin tensor term shown in the upper panel of Fig. 4(c) can be expressed as:

$$\sum_{ii'} p_{ii'}^{\gamma} = (1/6)^2 \sum_{ii'} P_{ii'}^{\alpha\beta\gamma} \mu_S^{\alpha} \mu_L^{\beta} = (1/6)^2 \mu_S^{\alpha} \mu_L^{\beta} (\sum_{ii'} P_{ii'}^{\alpha\beta\gamma}) \quad (6)$$

where $ii' = 11', 22'$ and $33'$. Similarly, after considering the $3_z$ and $m_y$ symmetries and applying Neumann's Principle, the sum of the two-spin tensors $\sum_{ii'} P_{ii'}^{\alpha\beta\gamma}$ between the two layers in the upper panel of Fig. 4(c) can be simplified as

$$\sum_{ii'} P_{ii'}^{\alpha\beta\gamma} = 3\begin{pmatrix} a',0,b' & 0,-a',0 & c',0,0 \\ 0,-a',0 & -a',0,b' & 0,c',0 \\ c'',0,0 & 0,c'',0 & 0,0,d' \end{pmatrix} \quad (7)$$

$$a' = \frac{P_{11'}^{xxx} - P_{11'}^{xyy} - P_{11'}^{yxy} - P_{11'}^{yyx}}{4}, b' = \frac{P_{11'}^{xxz} + P_{11'}^{yyz}}{2}, c' = \frac{P_{11'}^{xzx} + P_{11'}^{yzy}}{2}, c'' = \frac{P_{11'}^{zxx} + P_{11'}^{zyy}}{2}, d' = P_{11'}^{zzz}$$

where $P_{11'}^{\alpha\beta\gamma}$ are the matrix components of two-spin tensor of site-pair $11'$.

Moreover, there are extra symmetry operations between two layers within a magnetic block: space inversion for $S$ & $L$ blocks in Y-type hexaferrite and $S$ block in Z-type hexaferrite, mirror symmetry $m_z$ for $L$ block in Z-type hexaferrite, as shown in Fig. 4. The sum of the single-spin tensor $\sum_i P_{ii}^{\alpha\beta\gamma}$ at two layers in one block can be simplified further. If there is a space inversion at the block center, the sum of $\sum_i P_{ii}^{\alpha\beta\gamma}$ in the two layers of the block will be exactly opposite because they can be mutually transformed by **-1** symmetry. Then, the total summation matrix of $\sum_i P_{ii}^{\alpha\beta\gamma}$ within one block will be exactly zero for every component. Whereas, if there is a mirror in the middle of the block, the matrix of $\sum_i P_{ii}^{\alpha\beta\gamma}$ become:

$$\sum_i P_{ii}^{\alpha\beta\gamma} = 6\begin{pmatrix} a,0,0 & 0,-a,0 & 0,0,0 \\ 0,-a,0 & -a,0,0 & 0,0,0 \\ 0,0,0 & 0,0,0 & 0,0,0 \end{pmatrix} \tag{8}$$

The sum of the two-spin tensor $\sum_{ii'} P_{ii'}^{\alpha\beta\gamma}$ between two layers in one block can be also be simplified further. If there is a space inversion at the block center, as shown in the middle panel of Fig. 4(c), the matrix of $\sum_{ii'} P_{ii'}^{\alpha\beta\gamma}$ within one block will be simplified as:

$$\sum_{ii'} P_{ii'}^{\alpha\beta\gamma} = 3\begin{pmatrix} 0,0,0 & 0,0,0 & c',0,0 \\ 0,0,0 & 0,0,0 & 0,c',0 \\ -c',0,0 & 0,-c',0 & 0,0,0 \end{pmatrix} \tag{9}$$

Note that the number of spin-pair can be six in this case. However, this will not affect the number of independent coefficient, the form of matrix and the net polarization of two hexaferrite systems in Eqs. (8) and (11). While if there is an $m_z$ at the block center, as shown in the lower panel of Fig. 4(c), the matrix of $\sum_{ii'} P_{ii'}^{\alpha\beta\gamma}$ within one block will be simplified as:

$$\sum_{ii'} P_{ii'}^{\alpha\beta\gamma} = 3\begin{pmatrix} a',0,0 & 0,-a',0 & c',0,0 \\ 0,-a',0 & -a',0,0 & 0,c',0 \\ -c',0,0 & 0,-c',0 & 0,0,0 \end{pmatrix} \tag{10}$$

where the summation goes over every site $i$ or site pair $ii'$ in one block.

From the above procedures, we have greatly simplified the matrix forms of the summation of local ME tensors $\sum_i P_{ii}^{\alpha\beta\gamma}$ and $\sum_{ii'} P_{ii'}^{\alpha\beta\gamma}$ for both inter-block and intra-block cases. The forms of local ME tensor matrices for intra-block summations in $S_1$ and $L_1$ blocks, and inter-block summation between $S_1$ and $L_1$ block are determined in Tables 1 and 2. Then, those of other inter and intra-block summations in two hexaferrite systems can be derived by applying space inversion or $m_z$ symmetry operator according to Supplementary Table 1.

## Supplementary Note 3: THE SIMPLIFICATION OF SINGLE-SPIN TENSORS WITH WYCKOFF SITE SYMMETRY OPERATIONS

We could calculate the two-spin ME tensor matrices of $P_{12}^{\alpha\beta\gamma}$ for the 32 symmetry point groups From Neumann's Principle (Supplementary Eq. 2) and Supplementary Table 1, as summarized in Supplementary Table 2. The single-spin ME tensor matrices of $P_{11}^{\alpha\beta\gamma}$ for the various symmetry point groups can be deduced similarly.

Supplementary Table 2. The matrices of two-spin tensor $P_{12}^{\alpha\beta\gamma}$ in 32 point groups.

| | Simplified $P_{12}^{\alpha\beta\gamma}$ in matrix form |
|---|---|
| **1** | $\begin{pmatrix} P_{12}^{xx(x)}, P_{12}^{xx(y)}, P_{12}^{xx(z)} & P_{12}^{xy(x)}, P_{12}^{xy(y)}, P_{12}^{xy(z)} & P_{12}^{xz(x)}, P_{12}^{xz(y)}, P_{12}^{xz(z)} \\ P_{12}^{yx(x)}, P_{12}^{yx(y)}, P_{12}^{yx(z)} & P_{12}^{yy(x)}, P_{12}^{yy(y)}, P_{12}^{yy(z)} & P_{12}^{yz(x)}, P_{12}^{yz(y)}, P_{12}^{yz(z)} \\ P_{12}^{zx(x)}, P_{12}^{zx(y)}, P_{12}^{zx(z)} & P_{12}^{zy(x)}, P_{12}^{zy(y)}, P_{12}^{zy(z)} & P_{12}^{zz(x)}, P_{12}^{zz(y)}, P_{12}^{zz(z)} \end{pmatrix}$ |
| **2$_y$** | $\begin{pmatrix} 0, P_{12}^{xx(y)}, 0 & P_{12}^{xy(x)}, 0, P_{12}^{xy(z)} & 0, P_{12}^{xz(y)}, 0 \\ P_{12}^{yx(x)}, 0, P_{12}^{yx(z)} & 0, P_{12}^{yy(y)}, 0 & P_{12}^{yz(x)}, 0, P_{12}^{yz(z)} \\ 0, P_{12}^{zx(y)}, 0 & P_{12}^{zy(x)}, 0, P_{12}^{zy(z)} & 0, P_{12}^{zz(y)}, 0 \end{pmatrix}$ |
| **m$_y$** | $\begin{pmatrix} P_{12}^{xx(x)}, 0, P_{12}^{xx(z)} & 0, P_{12}^{xy(y)}, 0 & P_{12}^{xz(x)}, 0, P_{12}^{xz(z)} \\ 0, P_{12}^{yx(y)}, 0 & P_{12}^{yy(x)}, 0, P_{12}^{yy(z)} & 0, P_{12}^{yz(y)}, 0 \\ P_{12}^{zx(x)}, 0, P_{12}^{zx(z)} & 0, P_{12}^{zy(y)}, 0 & P_{12}^{zz(x)}, 0, P_{12}^{zz(z)} \end{pmatrix}$ |
| **mm2** | $\begin{pmatrix} 0, 0, P_{12}^{xx(z)} & 0, 0, 0 & P_{12}^{xz(x)}, 0, 0 \\ 0, 0, 0 & 0, 0, P_{12}^{yy(z)} & 0, P_{12}^{yz(y)}, 0 \\ P_{12}^{zx(x)}, 0, 0 & 0, P_{12}^{zy(y)}, 0 & 0, 0, P_{12}^{zz(z)} \end{pmatrix}$ |
| **222** | $\begin{pmatrix} 0, 0, 0 & 0, 0, P_{12}^{xy(z)} & 0, P_{12}^{xz(y)}, 0 \\ 0, 0, P_{12}^{yx(z)} & 0, 0, 0 & P_{12}^{yz(x)}, 0, 0 \\ 0, P_{12}^{zx(y)}, 0 & P_{12}^{zy(x)}, 0, 0 & 0, 0, 0 \end{pmatrix}$ |
| **3$_z$** | $\begin{pmatrix} P_{12}^{xx(x)}, P_{12}^{xx(y)}, P_{12}^{xx(z)} & P_{12}^{xx(y)}, -P_{12}^{xx(x)}, P_{12}^{xy(z)} & P_{12}^{xz(x)}, P_{12}^{xz(y)}, 0 \\ P_{12}^{xx(y)}, -P_{12}^{xx(x)}, -P_{12}^{xy(z)} & -P_{12}^{xx(x)}, -P_{12}^{xx(y)}, P_{12}^{xx(z)} & -P_{12}^{xz(y)}, P_{12}^{xz(x)}, 0 \\ P_{12}^{zx(x)}, P_{12}^{zx(y)}, 0 & -P_{12}^{zx(y)}, P_{12}^{zx(x)}, 0 & 0, 0, P_{12}^{zz(z)} \end{pmatrix}$ |
| $3_z2_x$ | $\begin{pmatrix} P_{12}^{xx(x)},0,0 & 0,-P_{12}^{xx(x)},P_{12}^{xy(z)} & 0,P_{12}^{xz(y)},0 \\ 0,-P_{12}^{xx(x)},-P_{12}^{xy(z)} & -P_{12}^{xx(x)},0,0 & -P_{12}^{xz(y)},0,0 \\ 0,P_{12}^{zx(y)},0 & -P_{12}^{zx(y)},0,0 & 0,0,0 \end{pmatrix}$ |
| $3_zm_y$ | $\begin{pmatrix} P_{12}^{xx(x)},0,P_{12}^{xx(z)} & 0,-P_{12}^{xx(x)},0 & P_{12}^{xz(x)},0,0 \\ 0,-P_{12}^{xx(x)},0 & -P_{12}^{xx(x)},0,P_{12}^{xx(z)} & 0,P_{12}^{xz(x)},0 \\ P_{12}^{zx(x)},0,0 & 0,P_{12}^{zx(x)},0 & 0,0,P_{12}^{zz(z)} \end{pmatrix}$ |
| $4_z$, $6_z$ | $\begin{pmatrix} 0,0,P_{12}^{xx(z)} & 0,0,P_{12}^{xy(z)} & P_{12}^{xz(x)},P_{12}^{xz(y)},0 \\ 0,0,-P_{12}^{xy(z)} & 0,0,P_{12}^{xx(z)} & -P_{12}^{xz(y)},P_{12}^{xz(x)},0 \\ P_{12}^{zx(x)},P_{12}^{zx(y)},0 & -P_{12}^{zx(y)},P_{12}^{zx(x)},0 & 0,0,P_{12}^{zz(z)} \end{pmatrix}$ |
| -4 | $\begin{pmatrix} 0,0,P_{12}^{xx(z)} & 0,0,P_{12}^{xy(z)} & P_{12}^{xz(x)},P_{12}^{xz(y)},0 \\ 0,0,P_{12}^{xy(z)} & 0,0,-P_{12}^{xx(z)} & P_{12}^{xz(y)},-P_{12}^{xz(x)},0 \\ P_{12}^{zx(x)},P_{12}^{zx(y)},0 & P_{12}^{zx(y)},-P_{12}^{zx(x)},0 & 0,0,0 \end{pmatrix}$ |
| 4mm, 6mm | $\begin{pmatrix} 0,0,P_{12}^{xx(z)} & 0,0,0 & P_{12}^{xz(x)},0,0 \\ 0,0,0 & 0,0,P_{12}^{xx(z)} & 0,P_{12}^{xz(x)},0 \\ P_{12}^{zx(x)},0,0 & 0,P_{12}^{zx(x)},0 & 0,0,P_{12}^{zz(z)} \end{pmatrix}$ |
| 422, 622 | $\begin{pmatrix} 0,0,0 & 0,0,P_{12}^{xy(z)} & 0,P_{12}^{xz(y)},0 \\ 0,0,-P_{12}^{xy(z)} & 0,0,0 & -P_{12}^{xz(y)},0,0 \\ 0,P_{12}^{zx(y)},0 & -P_{12}^{zx(y)},0,0 & 0,0,0 \end{pmatrix}$ |
| -42m | $\begin{pmatrix} 0,0,0 & 0,0,P_{12}^{xy(z)} & 0,P_{12}^{xz(y)},0 \\ 0,0,P_{12}^{xy(z)} & 0,0,0 & P_{12}^{xz(y)},0,0 \\ 0,P_{12}^{zx(y)},0 & P_{12}^{zx(y)},0,0 & 0,0,0 \end{pmatrix}$ |
| -6 | $\begin{pmatrix} P_{12}^{xx(x)},P_{12}^{xx(y)},0 & P_{12}^{xx(y)},-P_{12}^{xx(x)},0 & 0,0,0 \\ P_{12}^{xx(y)},-P_{12}^{xx(x)},0 & -P_{12}^{xx(x)},-P_{12}^{xx(y)},0 & 0,0,0 \\ 0,0,0 & 0,0,0 & 0,0,0 \end{pmatrix}$ |
| $-6m_y2$ | $\begin{pmatrix} P_{12}^{xx(x)},0,0 & 0,-P_{12}^{xx(x)},0 & 0,0,0 \\ 0,-P_{12}^{xx(x)},0 & -P_{12}^{xx(x)},0,0 & 0,0,0 \\ 0,0,0 & 0,0,0 & 0,0,0 \end{pmatrix}$ |
| -43m, 23 | $\begin{pmatrix} 0,0,0 & 0,0,P_{12}^{xy(z)} & 0,P_{12}^{xy(z)},0 \\ 0,0,P_{12}^{xy(z)} & 0,0,0 & P_{12}^{xy(z)},0,0 \\ 0,P_{12}^{xy(z)},0 & P_{12}^{xy(z)},0,0 & 0,0,0 \end{pmatrix}$ |
| 432 | $\begin{pmatrix} 0,0,0 & 0,0,P_{12}^{xy(z)} & 0,-P_{12}^{xy(z)},0 \\ 0,0,-P_{12}^{xy(z)} & 0,0,0 & P_{12}^{xy(z)},0,0 \\ 0,P_{12}^{xy(z)},0 & -P_{12}^{xy(z)},0,0 & 0,0,0 \end{pmatrix}$ |
| **others** | $\begin{pmatrix} 0,0,0 & 0,0,0 & 0,0,0 \\ 0,0,0 & 0,0,0 & 0,0,0 \\ 0,0,0 & 0,0,0 & 0,0,0 \end{pmatrix}$ |

From the Supplementary Table 2, we could deduce the simplified single-spin ME tensor matrices of $P_{11}^{\alpha\beta\gamma}$ for 3$m$, $m$ and -6$m$2 respectively in Supplementary Table 3:

Supplementary Table 3. The matrices of single-spin tensor $P_{11}^{\alpha\beta\gamma}$ in selected point groups

| | Simplified matrices of $P_{11}^{\alpha\beta\gamma}$ |
|---|---|
| $m$ | $\begin{pmatrix} P_{11}^{xx(x)},0,P_{11}^{xx(z)} & 0,P_{11}^{xy(y)},0 & P_{11}^{xz(x)},0,P_{11}^{xz(z)} \\ 0,P_{11}^{xy(y)},0 & P_{11}^{yy(x)},0,P_{11}^{yy(z)} & 0,P_{11}^{yz(y)},0 \\ P_{11}^{xz(x)},0,P_{11}^{xz(z)} & 0,P_{11}^{yz(y)},0 & P_{11}^{zz(x)},0,P_{11}^{zz(z)} \end{pmatrix}$ |
| 3$m$ | $\begin{pmatrix} P_{11}^{xx(x)},0,P_{11}^{xx(z)} & 0,-P_{11}^{xx(x)},0 & P_{11}^{xz(x)},0,0 \\ 0,-P_{11}^{xx(x)},0 & -P_{11}^{xx(x)},0,P_{11}^{xx(z)} & 0,P_{11}^{xz(x)},0 \\ P_{11}^{xz(x)},0,0 & 0,P_{11}^{xz(x)},0 & 0,0,P_{11}^{zz(z)} \end{pmatrix}$ |
| -6$m$2 | $\begin{pmatrix} P_{11}^{xx(x)},0,0 & 0,-P_{11}^{xx(x)},0 & 0,0,0 \\ 0,-P_{11}^{xx(x)},0 & -P_{11}^{xx(x)},0,0 & 0,0,0 \\ 0,0,0 & 0,0,0 & 0,0,0 \end{pmatrix}$ |

From the site symmetries shown in Table 5, all the site symmetries at Me3 to Me10 fall in the above three symmetries. According to Supplementary Table 3, those sites should have non-zero $a_0$ in Eq. 9 if there is no extra or hidden symmetry constraints.